\documentstyle[epsf,multicol,aps]{revtex}
\begin{document}
\draft 

\title{Supernova electron capture rates for $^{55}$Co and
  $^{56}$Ni} 

\author{K. Langanke and G. Mart\'{\i}nez-Pinedo} 

\address{Institute for Physics and Astronomy, University of {\AA}rhus,
  Denmark and Theoretical Center for Astrophysics, University of
  {\AA}rhus, Denmark} 

\date{\today} 

\maketitle

\begin{abstract}
  We have calculated the Gamow-Teller strength distributions for the
  ground states and first excited states in $^{55}$Co and $^{56}$Ni.
  These calculations have been performed by shell model
  diagonalization in the $pf$ shell using the KB3 interaction.  The
  Gamow-Teller distributions are used to calculate the electron
  capture rates for typical presupernova conditions. Our $^{55}$Co
  rate is noticeably smaller than the presently adopted rate as it is
  dominated by weak low-lying transitions rather than the strong
  Gamow-Teller (GT) resonance which is located at a higher excitation
  energy in the daughter than usually parametrized. Although our
  $^{56}$Ni rate agrees with the presently adopted rate, we do not
  confirm the conventional parametrization of the GT centroid.  Our
  results support general trends suggested on the basis of shell model
  Monte Carlo calculations.
\end{abstract}


\begin{multicols}{2}
  
\section{Introduction}

The core of a massive star becomes dynamically unstable when it
exhausts its nuclear fuel. If the core mass exceeds the appropriate
Chandrasekhar mass, electron degeneracy pressure cannot longer
stabilize the center and it collapses. As pointed out by Bethe {\it et
  al.}  \cite{Bethe78,Bethe90} the collapse is very sensitive to the
entropy and to the number of leptons per baryon, $Y_e$. In the early
stage of the collapse $Y_e$ is reduced as electrons are captured by Fe
peak nuclei.  Knowing the importance of the electron capture process,
Fuller {\it et al.} (usually called FFN) have systematically estimated
the rates for nuclei in the mass range $A=45-60$ putting special
emphasis on the importance of capture to the Gamow-Teller (GT) giant
resonance \cite{FFN}.  The GT contribution to the rate has been
parametrized by FFN on the basis of the independent particle model. To
complete the FFN rate estimate, the GT contribution has been
supplemented by a contribution simulating low-lying transitions.
Recently the FFN rates have been updated and extended to heavier
nuclei by Aufderheide {\it et al.} \cite{Aufderheide}.  These authors
also considered the wellknown quenching of the Gamow-Teller strength
by reducing the independent particle estimate for the GT resonance
contribution by a common factor of two.

After experimental (n,p) data clearly indicated that the Gamow-Teller
strength is not only quenched (usually by more than a factor 2
compared to the independent particle model), but also fragmented over
several states at modest excitation energies in the daughter nucleus
\cite{gtdata1,gtdata2,gtdata3,gtdata4,gtdata5}, the need for an
improved theoretical description has soon been realized
\cite{Aufderheide91,Aufderheide93a,Aufderheide93b}.  These studies
have been performed within the conventional shell model
diagonalization approach, however, in strongly restricted model spaces
and with residual interactions, which turned out to neither reproduce
the quenching nor the position of the GT strength sufficiently well.
These model studies therefore had only a limited value, as they
required experimental input informations, and they had no predictive
power. This situation changed recently as the development of the shell
model Monte Carlo technique (SMMC) \cite{Johnson92,Dean94,Physrep}
allows calculations of the GT strength distribution in the complete
$pf$ shell. In fact, using the KB3 residual interaction \cite{Zuker}
it has been demonstrated that both the quenching of the GT strength
\cite{Langanke95} and its distribution \cite{Radha97} can be well
reproduced. Using the SMMC method, Dean {\it et al.} have recently
calculated electron capture rates for several Fe peak nuclei of
importance at the early stage of the presupernova collapse
\cite{Dean97}.  This calculation indicated systematic differences in
the location of the main GT resonance strength compared to the
parametrization of FFN.  In capture on even-even nuclei the GT
strength resides at lower excitation energies in the daughter than
assumed by FFN, while in odd-A nuclei the GT strength is centered at
higher excitation energies. The same trend is also seen in the
available (n,p) data \cite{Koonin94} and has been pointed out for
individual cases in \cite{Aufderheide93a,Aufderheide93b}.

Ref. \cite{Dean97} demonstrates that, for even-even parent nuclei, the
electron capture rates are given by the bulk of the Gamow-Teller
strength distribution, which resides at low excitation energies in the
daughter, and is well reproduced by the SMMC method. The situation is
quite different in odd-A nuclei. Here the bulk of the GT strength is
at a too high excitation energy to be of significance for the electron
capture rates which are dominated by weak low-lying transitions.
Unfortunately the SMMC method is not capable of spectroscopy and does
not allow to extract these weak transition strengths. These
informations, however, can be obtained from shell model
diagonalizations techniques which have made significant progress in
the last couple of years to allow now for basically complete $pf$
shell diagonalizations for low-lying states in the $A=56$ mass range
\cite{Nowacki}.

Aufderheide {\it et al.} have ranked the core nuclei with respect to
their importance for the electron capture process in the presupernova
\cite{Aufderheide}.  As the two most important nuclei these authors
identified $^{55}$Co and $^{56}$Ni for the early presupernova
collapse. Both rates, however, should be strongly affected by the
misplacement of the GT resonance position. A new estimate for the
capture rate on $^{56}$Ni has already been given in \cite{Dean97}
based on an SMMC calculation, however, this rate has been cautioned by
the authors due to possible oberbinding effects at the $N=28$ shell
closure in their approach.  Due to their importance a calculation of
the two rates on the basis of a shell model diagonalization approach
seems to be quite useful.  We have performed such a calculation for
the $^{56}$Ni ground state and the 3 lowest states in $^{55}$Co using
the KB3 interaction \cite{Zuker} and making use of the
state-of-the-art diagonalization code ANTOINE \cite{Caurier}.

It is wellknown that $0\hbar\omega$ shell model calculations, i.e.
calculations performed in one major shell, overestimate the GT
strength by a universal factor $(1.26)^2$
\cite{Wildenthal,Langanke95,Martinez96}, often interpreted as a
renormalization of the axialvector coupling constant $g_A $ in nuclei.
To account for this fact, we have used the renormalized value $g_A=1$
in the following..

For $^{56}$Ni we have calculated the total GT strength in a full $pf$
shell calculation, resulting in B(GT)=$g_A^2 | \langle {\vec \sigma}
\tau_+ \rangle |^2=10.1 g_A^2$ (see also \cite{Nowacki}), which is in
agreement with the SMMC value (B(GT)=$(9.8\pm0.4) g_A^2$
\cite{Langanke95}). The independent particle model yields B(GT)=$13.7
g_A^2$.  The GT strength distribution has been calculated in a model
space which allowed a maximum of 6 particles to be excited from the
$f_{7/2}$ shell to the rest of the $pf$-shell in the final nucleus,
$^{56}$Co. The $m$-scheme dimension of this calculation is 19831538.
In this truncated calculation we obtain a total GT strength of $10.2
g_A^2$, indicating that our calculation is almost converged at this
truncation level (see also \cite{iron}).  For $^{55}$Co we have
calculated the total GT strength and the distribution in a truncated
calculation which fulfills the Ikeda sum rule and in which maximally 5
particles in the final nucleus are allowed to be excited out of the
$f_{7/2}$ orbital. We obtain a total GT strength of $8.7 g_A^2$ from
the ground state of $^{55}$Co, and $8.9 g_A^2$ from both of the
excited $J=3/2$ states.  The values are to be compared with the
independent particle value of B(GT)=$12 g_A^2$.  We note that for
both, $^{55}$Co and $^{56}$Ni, the quenching factor is unusually small
due to the shell closure at $^{56}$Ni.

We have performed 33 Lanczos iterations which are usually sufficient
to converge in the states at excitation energies below $E= 3$ MeV.  At
higher excitation energies, $E>3$ MeV, the calculated GT strengths
represent centroids of strengths, which in reality are splitted over
many states.  For calculating the electron capture rate, however, a
resolution of this strength at higher energies is unimportant.

The GT strength distributions $S_{GT}(E)$ for the lowest states in
$^{55}$Co ($J=7/2$ ground state and the two $J=3/2$ excited states at
$E=2.165$ MeV and 2.565 MeV, respectively) and for the $^{56}$Ni
ground state are shown in Figs. 1 and 2. The energy scale in these
figures has been adjusted such that the lowest calculated state of a
given angular momentum agrees with the experimentally known excitation
energy; the necessary energy shifts have been less than 300 keV as the
shell model calculations reproduce the low-lying spectrum in $^{55}$Fe
and $^{56}$Co rather well.  We observe that for the $^{55}$Co ground
state the GT centroid resides at about $E=6$ MeV in the daughter
$^{55}$Fe, while it is at around $E = 9-10$ MeV for the excited
states. This result is in agreement with the SMMC study (performed at
temperature $T=0.8$ MeV) which found the centroid of the GT strength
at $E=6.9$ MeV in $^{55}$Fe \cite{Dean97}.

To estimate the electron capture rates at finite temperatures, the
compilations employed the so-called Brink hypothesis
\cite{Aufderheide,Aufderheide91}- assuming that the GT strength
distribution on excited states is the same as for the ground state,
only shifted by the excitation energy of the state.  For $^{55}$Co
this assumption is roughly valid for the bulk of the strength, but it
is clearly not justified for the low-lying transitions which are
dominated by the individual structures of the states involved.  While
the $^{55}$Co ground state has rather weak GT transitions to low-lying
states in $^{55}$Fe, the first excited $J=3/2$ state has a strong
transition to the $^{55}$Fe ground state (and first excited state).
The quality of our calculation can be tested by calculating the
lifetime of $^{55}$Co under terrestrial conditions where it decays by
$\beta_+$-decay.  Using the GT matrix elements as calculated in our
shell model approach and the experimental energy splittings we
calculate a $^{55}$Co lifetime of 16.7 hours, which compares nicely
with the experimental value of 17.53 hours.

Under presupernova conditions the electron capture on $^{56}$Ni is
dominated by the ground state as the first excited state is too high
in excitation energy.  The centroid of the GT strength is around
$E=2.5-3$ MeV in $^{56}$Co, in agreement with the SMMC estimate given
in Ref.  \cite{Dean97}. For comparison, FFN placed the GT resonance in
$^{56}$Co at E=3.8 MeV. For the lifetime of $^{56}$Ni we find 6.7 d,
very close to the experimental value of 6.08 d.

The presupernova electron capture rate $\lambda_{\rm ec}$ is given
by~\cite{FFN,Aufderheide}

\begin{eqnarray}
\lambda_{\rm{ec}} & = &\frac{\ln 2}{6163 \rm{sec}}
\sum_{ij} \frac{(2J_i+1) \exp{[-E_i/kT]}}{G}
S^{ij}_{\rm{GT}} 
\frac{c^3}{(m_e c^2)^5}\nonumber\\
&&\int_{\cal L}^\infty dp p^2 (Q_{ij}+E_e)^2
\frac{F(Z,E_e)}{1+\exp\left[\beta_e(E_e-\mu_e)\right]}\;,
\end{eqnarray}
where $E_e$, $p$, and $\mu_e$ are the electron energy, momentum, and
chemical potential, and ${\cal L}=(Q_{if}^2 - m_e^2 c^4)^{1/2}$ for
$Q_{if} \leq -m_e c^2$, and 0 otherwise. $Q_{if}=E_i-E_f$ is the
nuclear energy difference between the initial and final states, while
$S^{ij}_{\rm {GT}}$ is their GT transition strength.  G is the
partition function, $G=\sum_i (2J_i+1) \exp{[-E_i/kT]}$.  The Fermi
function $F(Z,E_e)$ accounts for the distortion of the electron's wave
function due to the Coulomb field of the nucleus.

The calculated electron capture rates for $^{55}$Co and $^{56}$Ni are
shown in Figs. 3 and 4 as function of temperature ($T_9 $ measures the
temperature in $10^9$ K) and for selected densities ($\rho_7$ measures
the density in $10^7$ g/cm$^3$). For the chemical potential we use the
approximation \cite{Aufderheide90}
\begin{equation}
\mu_e = 1.11 (\rho_7 Y_e)^{1/3}\left[1+\left(\frac{\pi}{1.11}\right)^2
\frac{T^2}{\left(\rho_7 Y_e\right)^{2/3}}\right]^{-1/3}\;.
\end{equation}
Due to the ranking given in Ref. \cite{Aufderheide}, $^{55}$Co and
$^{56}$Ni are the most important electron capture nuclei at
temperatures and densities around $T_9=3.26$ and $\rho_7=4.32$. Under
these conditions the recommended capture rates (the FFN rates are in
parenthesis) for $^{55}$Co and $^{56}$Ni are \cite{Aufderheide}
$\lambda$= $5.1 \cdot 10^{-2}$ s$^{-1}$ ($8.4 \cdot 10^{-2}$ s$^{-1}$)
and $8.6 \cdot 10^{-3}$ s$^{-1}$ ($7.5 \cdot 10^{-3}$ s$^{-1}$),
respectively, while our calculation yields $1.6 \cdot 10^{-3}$
s$^{-1}$ for $^{55}$Co and $12.6 \cdot 10^{-3}$ s$^{-1}$ for
$^{56}$Ni. For $^{56}$Ni the three rates agree rather well, but this
agreement is more or less accidental as a closer inspection shows. In
the calculation of \cite{Aufderheide} the dominant contribution comes
from the transition to the low-lying states (simulated for all nuclei
by an effective B(GT)=0.1 for a transition to a fictitious state at
$E=0$), while only $18\%$ originates from the GT resonance placed at
$E=3.8$ MeV. We, however, find that nearly $50\%$ of the capture rate
is due to the strong transition to the GT resonance, which in our
calculation is located nearly 1 MeV lower in excitation energy than
parametrized in Refs.  \cite{FFN,Aufderheide}. Our $^{56}$Ni rate also
approximately agrees with the SMMC estimate of Ref.  \cite{Dean97}.

As already suggested in \cite{Dean97} the recommended rate for
$^{55}$Co is too large (by more than an order of magnitude), as the
authors of \cite{FFN,Aufderheide} placed the GT resonance at too low
an excitation energy and consequently assigned 73$\%$ and 83$\%$,
respectively, to this transition. In contrast, our calculated rate is
predominantly given by transitions to the low-lying states. In fact,
we recover about 80$\%$ of the rate if we cut the calculated GT
spectra at the excitation energy $E=3$ MeV. We also note that the
$^{55}$Co rate arises mainly from capture on the ground state at
temperatures $T_9\approx3.3$ (which are typical for presupernova
electron capture on $^{55}$Co), while, due to the rather large
excitation energies, contributions from capture on the excited states
amount to less than $5\%$ under the relevant presupernova conditions.

In summary, we have performed state-of-the-art large-scale shell model
diagonalization calculations to determine the presupernova electron
capture rates on $^{55}$Co and $^{56}$Ni, which are believed to be the
most important ``electron poisons'' at the onset of collapse. Although
our calculation approximately agrees with the recommended rate for
$^{56}$Ni, it does not confirm the parametrization conventionally used
to derive at these rates. Our calculation finds the bulk of the GT
strength distribution at around 1 MeV lower in excitation energy than
assumed in the parametrization. As has already been noted before
\cite{Aufderheide93a,Aufderheide93b,Dean97} this trend seems to be
general for even-even parent nuclei, while for odd-A nuclei the
parametrization places the GT centroid at too low excitation energies.
This suggestion \cite{Dean97} is confirmed in our shell model
calculation. In fact we find that the centroid is too high in
excitation energy to affect noticeably the electron capture rate under
presupernova conditions.  Consequently our calculated rate is more
than one order of magnitude smaller than the compiled rates
\cite{FFN,Aufderheide}.

What are the consequences for the presupernova collapse? At the onset
of collapse ($\rho_7 \approx4.3$) the change of $Y_e$ with
time,$dY_e/dt$, has been assumed to be predominantly due to electron
capture on $^{55}$Co ($50\%$) and $^{56}$Ni ($25\%$)
\cite{Aufderheide}. With our revised rates, $^{56}$Ni becomes the
dominant source for electron capture and $dY_e/dt$ is reduced by
nearly a factor of 2.  However, for firm conclusions it appears to be
reasonable to first update the electron capture rates on all important
nuclei and then to perform a simulation of the presupernova collapse.
Such a program is in progress.

\acknowledgements

Discussions with David Dean are gratefully acknowledged.  This work
was supported in part by the Danish Research Council.  Grants of
computational resources were provided by the Center for Advanced
Computational Research at Caltech.

\end{multicols}

\begin{multicols}{2}
\narrowtext

\begin{figure}
  \begin{center}
    \epsfxsize=0.45\textwidth
    \epsffile{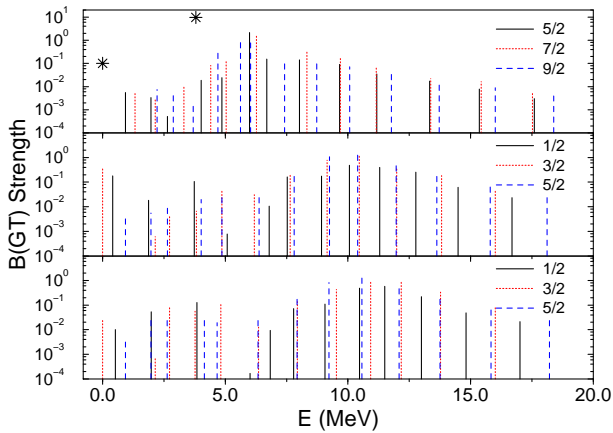}
    \caption{Gamow-Teller strength distributions for the $^{55}$Co
      ground state (top panel), the $J=3/2$ excited state at $E=2.165$
      MeV (middle panel) and the $J=3/2$ excited state at $E=2.565$
      MeV (lower panel).  For comparison the parametrized GT spectrum
      assumed in \protect\cite{Aufderheide} is indicated by stars in
      the top panel.  This parametrization assumed a fictitious state
      at $E=0$ and the GT resonance with a strength half of the
      independent particle model value.  The energy scale refers to
      excitation energies in the daughter nucleus, where we have
      shifted the calculated energies as to match the lowest
      experimentally known for a given angular momentum.}
    \label{fig1}
  \end{center}
\end{figure}

\begin{figure}
  \begin{center}
    \epsfxsize=0.45\textwidth
    \epsffile{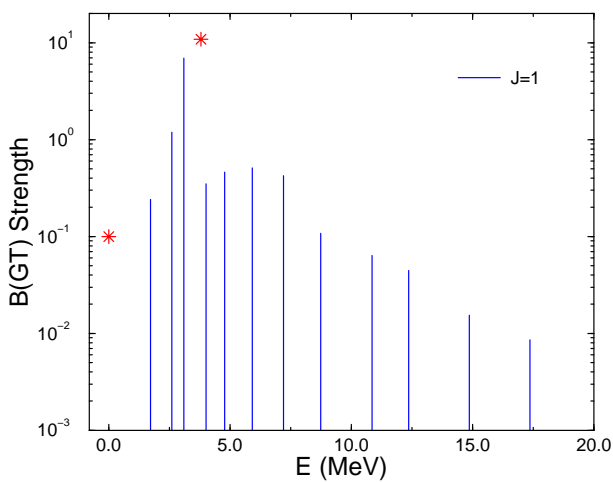}
    \caption{Gamow-Teller strength distribution for the $^{56}$Ni ground
      state.  For comparison the parametrized GT spectrum assumed in
      \protect\cite{Aufderheide} is indicated by stars (see Fig. 1).
      The energy scale refers to excitation energies in the daughter
      nucleus, where we have shifted the calculated energies as to
      match the lowest experimentally known $J=1$ state in $^{56}$Co.}
    \label{fig2}    
  \end{center}
\end{figure}

\begin{figure}
  \begin{center}
    \epsfxsize=0.45\textwidth
    \epsffile{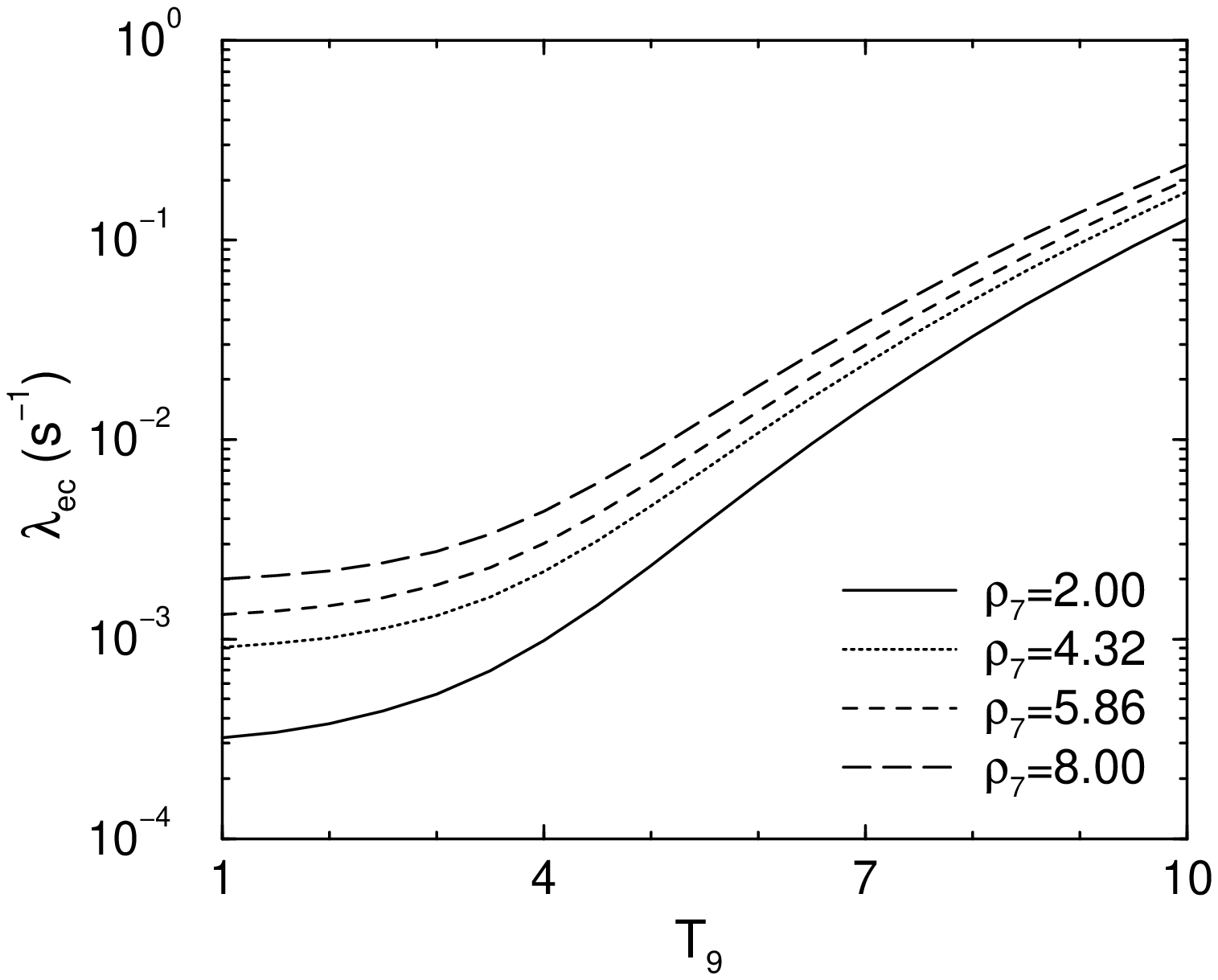}
    \caption{Electron capture rates on $^{55}$Co as function of
      temperature and for selected densities.}
    \label{fig3}
  \end{center}
\end{figure}

\begin{figure}
  \begin{center}
    \epsfxsize=0.45\textwidth
    \epsffile{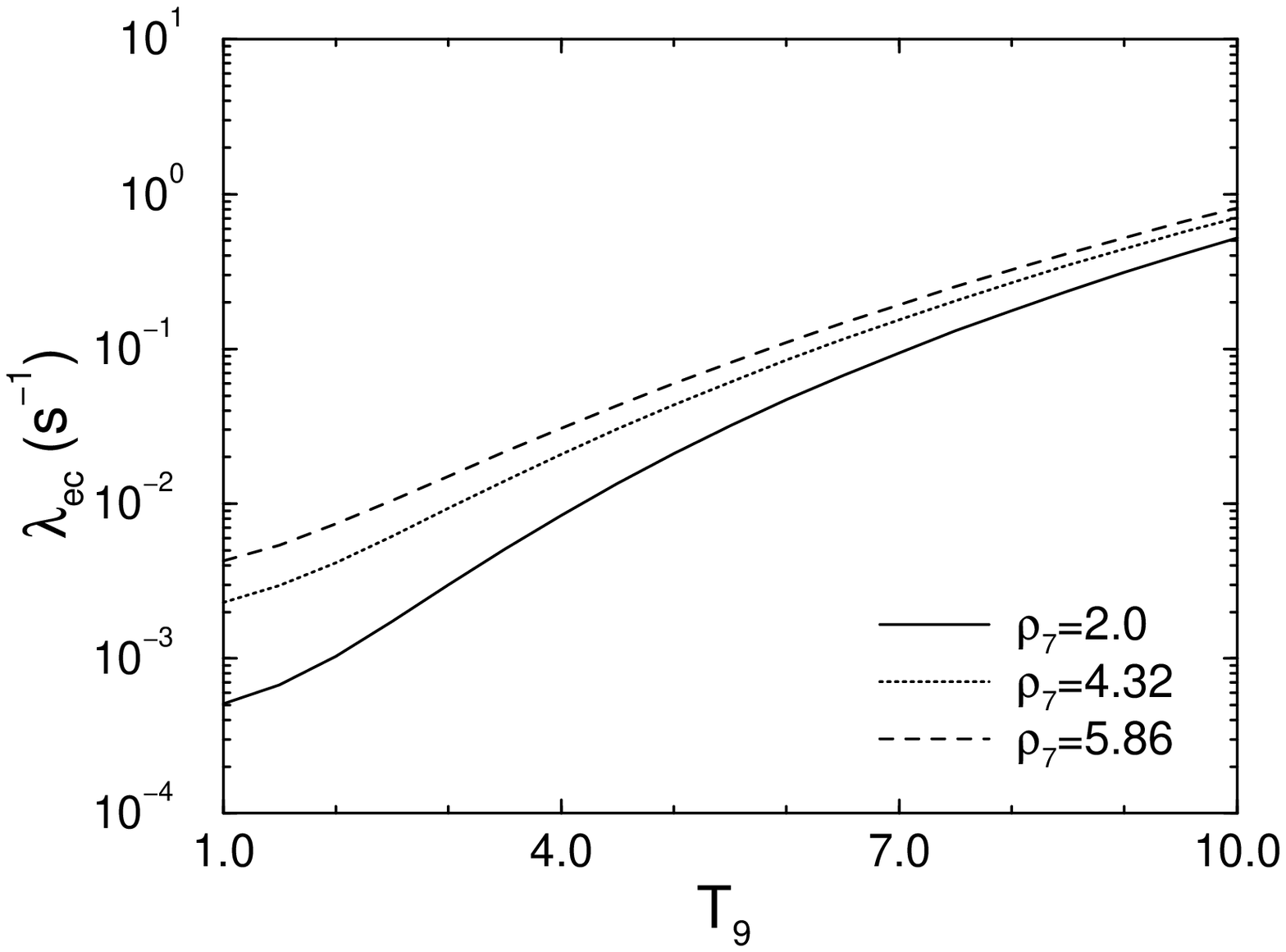}
    \caption{Electron capture rates on $^{56}$Ni as function of
      temperature and for selected densities.}
    \label{fig4}
  \end{center}
\end{figure}

\end{multicols}

\end{document}